\documentclass[12pt,preprint,aps,prd]{revtex4-1}
\usepackage{amsmath,amssymb,mathrsfs,graphicx,slashed,ragged2e}

\usepackage{hyperref}
\usepackage{float}
\usepackage{graphicx}
\usepackage{dcolumn}
\usepackage{bm}

\DeclareMathOperator{\Tr}{Tr}

\newcommand{\mM}{\text{Met}(M)}

\newcommand{\rH}{r_{\mathcal{H}}}

\begin{document}

\title[Black hole superspaces]{Quantifying closeness between black hole spacetimes: a superspace approach}

\author{Arthur George Suvorov}
\email{arthur.suvorov@tat.uni-tuebingen.de}
\affiliation{Theoretical Astrophysics, IAAT, University of T{\"u}bingen, T{\"u}bingen 72076, Germany}

\date{\today}

\begin{abstract}
\noindent{The set of all metrics that can be placed on a given manifold defines an infinite-dimensional `superspace' that can itself be imbued with the structure of a Riemannian manifold. Geodesic distances between points on $\text{Met}(M)$ measure how close two different metrics over $M$ are to one another. Restricting our attention to only those metrics that describe physical black holes, these distances may therefore be thought of as measuring the level of geometric similarity between different black hole structures. This allows for a systematic quantification of the extent to which a black hole, possibly arising as an exact solution to a theory of gravity extending general relativity in some way, might be `non-Kerr'. In this paper, a detailed construction of a superspace for stationary black holes with an arbitrary number of hairs is carried out. As an example application, we are able to strengthen a recent claim made by Konoplya and Zhidenko about which deviation parameters describing a hypothetical, non-Schwarzschild black hole are likely to be most relevant for astrophysical observables.}
\end{abstract}


\maketitle

\section{Introduction}

The study of geometrodynamics championed by Wheeler \cite{wheeler1,wheeler2,wheeler3} aims to explore, in some precise sense, the configuration space of general relativity (GR). Remarkably, the collection of all metrics that can be placed on some manifold admits a rich geometric structure: the set $\mM$ of metrics that can be placed over $M$ can itself be viewed as an (infinite-dimensional) Riemannian manifold \cite{gil2,clark10,ebin,gil1}. Considering a subspace of $\mM$ which includes only those metrics which are Einstein defines Wheeler's \emph{superspace} \cite{fishcer,edwards,guil09}, and provides a means to quantify the relationship between different spacetime structures in a purely geometrical setting. In particular, geodesics on superspace can be used to `measure' a distance between distinct metrics. More ambitiously, understanding the configuration space of GR may provide a stepping stone to the building of a quantum path integral over spacetime histories \cite{hawk,eucqg}.

In GR, all stationary, vacuum, and asymptotically flat black holes (BHs) are locally isometric to the Kerr solution \cite{kerr63,mars1}. This result is related to the no-hair theorems \cite{chand1}, and implies that an appropriately defined configuration space of BHs consists only of various Kerr metrics with different mass and spin parameters (see also Ref. \cite{schconfig}). However, if GR is in some sense incomplete and provides but an inexact description for the gravitational field in the strong-field regime, it is possible that more general BHs, described by some finite set of generalised charges or `hairs' $q^{i}$ (e.g. \cite{exact1,exact2,exact3,exact4}), may exist in Nature. In this paper, we note that this vector of hairs $\boldsymbol{q}$ can be used to introduce a coordinate basis over a generalised BH superspace $\text{Met}_{\text{BH}}(M)$. Some technical considerations necessary to build the superspace, such as the extraction of a single Riemannian manifold $M$ from a collection of BH spacetimes, are resolved in detail. Distances on this space, essentially quantifying the difference between two different hairy BHs, can then be used for benchmarking `non-Kerrness'; see also Ref. \cite{bhag18}. 
 
In general, the nonlinear and (often) higher-than-second order nature of non-GR field equations renders the task of finding exact solutions challenging, and only a handful of physically relevant ones are known (e.g. \cite{exact1,exact2,exact3,exact4}). To address the dearth of exact solutions, various approaches to constructing metrics representing generic BHs in a theory-agnostic manner have been developed \cite{nkerr1,nkerr2,nkerr3,contfrac}. These metrics are designed to represent parameterised departures from a Kerr description in some particular way, such as including deformations that still preserve the Killing tensor symmetry \cite{nkerr5,nkerr4}. Recently, Konoplya and Zhidenko (KZ hereafter) \cite{kz20} derived a metric which contains ``the only parameters that matter'', in the sense that they found including higher-order terms beyond those in their parameterised metric affected electromagnetic observables very little. As an application of the formalism developed here, we further validate the claims made in the aforementioned study by showing that including additional, non-KZ parameters, leads to relatively small curvature differentials on the BH superspace.

This paper is organised as follows. In Section 2 we introduce the concept of a relativistic superspace, and describe how it may be applied to the case of BH geometries. Section 3 then details the necessary technical aspects of the construction (Secs. 3.1 and 3.2), and walks the reader through the various steps for the simple case of static, spherically symmetric BHs (Sec. 3.3). Section 4 is devoted to some worked examples, namely the GR case of the Kerr \cite{kerr63} metric (Sec. 4.1) and the theory-agnostic KZ \cite{kz20} solution (Sec. 4.2). {Some general remarks about the applicability, and limitations, of the approach as a measure for BH closeness are detailed in Section 5.} Some discussion is given in Section 6.

\section{Superspaces}

The collection of all metrics definable over a Riemannian manifold $M$ can be shown to admit enough structure that it itself defines an (infinite-dimensional) Riemannian manifold \cite{ebin}. Points on $\mM$ are Riemannian metrics on $M$, i.e. each $p \in \mM$ corresponds to a symmetric, positive-definite $(0,2)$-tensor over $M$. We take the base-space $M$ to be 3-dimensional (though a generalisation to the higher-dimensional case is straightforward), as later it will be identified with the leaves of a spacelike foliation of a 4-dimensional BH spacetime.

A metric $G$ can be introduced over $\mM$, in the $L^{2}$-topology \cite{clark10}, as \cite{ebin,gil1}
\begin{equation} \label{eq:gilmed}
G(\mu,\nu) = \int_M d^{3} x \sqrt{h} \Tr \left( h^{-1} \mu h^{-1} \nu \right) ,
\end{equation}
where $\mu$ and $\nu$ are tangent vectors to the space of metrics at the `reference point' $h$, provided that the integral converges (see Sec. 3). It is important to note that other metrics exist on $\mM$, all of which fall under the class of the so-called $\alpha$-metrics \cite{osmo} which include \eqref{eq:gilmed} and the DeWitt \cite{gil2} metric as special cases. However, the choice \eqref{eq:gilmed} is invariant under the action of the diffeomorphism group $\text{Diff}(M)$ on $\mM$ \cite{gil1}, and is therefore `canonical' in some appropriate sense. Nevertheless, using a different metric (e.g. DeWitt \cite{gil2}), does not qualitatively change the picture discussed herein.

As it stands, the metric \eqref{eq:gilmed} is defined over the entire manifold $\mM$. In this work we are exclusively interested in those metrics that could describe BHs, so we restrict our consideration to only those $\mu$ and $\nu$ which are tangent vectors to some appropriately defined (see below) space of \emph{black hole metrics}, $\text{Met}_{\text{BH}}(M)$. The submanifold $\text{Met}_{\text{BH}}(M) \subset \mM$ inherits a metric from its parent space via pullback, which we also call $G$ through a slight abuse of notation. In theories other than GR, it is difficult to define $\text{Met}_{\text{BH}}(M)$ in total generality since, depending on the validity of the no-hair theorem \cite{chand1}, there may be an arbitrarily large (but finite) number of parameters (`hairs') which characterise the BH (cf. Refs. \cite{pappas15,suvmel16}). Nevertheless, suppose that a BH can be described by $N$ macroscopic hairs $q^{1}, \ldots, q^{N}$. The parameters $\boldsymbol{q}$ can be used to define a natural coordinate basis for the $N$-dimensional submanifold $\text{Met}_{\text{BH}}(M)$. In vacuum GR, Kerr uniqueness \cite{mars1} implies that $\text{Met}_{\text{GR-BH}}(M)$ is a two-dimensional space, spanned by appropriate mass ($q^{1}$, say) and spin ($q^{2}$, say) vectors. Electric \cite{chand1} or more general Yang-Mills \cite{ymbh1,ymbh2} charges giving rise to additional hairs may also be relevant for electrovacuum BHs in GR.

With respect to the $\boldsymbol{q}$ basis, the tensor components of \eqref{eq:gilmed} read \cite{suv20}
\begin{equation} \label{eq:metten}
G_{ij} = \int_{M} d^{3} x \sqrt{h}  h^{nk} \frac { \partial h_{mn}} {\partial q^{i}}   h^{\ell m}  \frac { \partial h_{\ell k}} {\partial q^{j}}  ,
\end{equation}
where $1 \leq i,j \leq N$. Assuming a (Levi-Civita) connection, all relevant geometric quantities on $\text{Met}_{\text{BH}}(M)$ can be defined from \eqref{eq:metten}, including the Christoffel symbols $\Gamma$. The distance between two metrics $\ell$ and $k$, described by parameter vectors $\boldsymbol{q}_{\ell}$ and $\boldsymbol{q}_{k}$, respectively, is then given by the length of a geodesic $\gamma: [0,1] \mapsto \text{Met}_{\text{BH}}(M)$ connecting these points, viz.
\begin{equation} \label{eq:distance}
d(\ell,k) = \int^{1}_{0} d \lambda \sqrt{ G_{ij} \frac {d \gamma^{i}} {d \lambda} \frac {d \gamma^{j}} {d \lambda} },
\end{equation}
for affine parameter $\lambda$, where $\gamma^{j}(0) = q_{\ell}^{j}$ and $\gamma^{j}(1) = q_{k}^{j}$, and $\gamma$ satisfies the geodesic equation,
\begin{equation} \label{eq:geodesic}
0 = \frac {d^2 \gamma^{i}} {d \lambda^2} + \Gamma^{i}_{j k} \frac {d \gamma^{j}} {d \lambda} \frac {d \gamma^{k}} {d \lambda}.
\end{equation}
From the Christoffel symbols $\Gamma$, we can also calculate the Ricci curvature
\begin{equation} \label{eq:ric}
R_{ij} = \Gamma^{k}_{ij,k} - \Gamma^{k}_{i k,j} + \Gamma^{k}_{ij} \Gamma^{\ell}_{\ell k} - \Gamma^{k}_{i \ell} \Gamma^{\ell}_{j k},
\end{equation}
whose contraction, $R = G^{ij} R_{ij}$, contains additional information about the structure of $\text{Met}_{\text{BH}}(M)$; see Sec. 4.

\section{Stationary, asymptotically flat black holes and 3+1 decompositions}


In GR and its extended variants, dynamics traditionally take place on a Lorentzian \emph{spacetime} $(\mathcal{M},g)$ rather than on a Riemannian manifold, i.e. the metric $g$ is not positive-definite. To construct the space $\text{Met}_{\text{BH}}(M)$ we must first extract a Riemannian $(\text{sub-})$manifold $M$ from the 4-dimensional pseudo-Riemannian manifold $\mathcal{M}$. The standard approach to achieving this is through a $3+1$ Arnowitt-Deser-Misner (ADM) split \cite{admsplit}; for a globally hyperbolic spacetime, a foliation of $\mathcal{M}$ by spacelike (Cauchy) hypersurfaces $\Sigma_{t}$ parameterised by some global time coordinate $t$ is always possible \cite{ger70}. For stationary spacetimes, the underlying leaves of the foliation are necessarily time-independent in some appropriate sense and we need only consider one such $\Sigma_{t}$, which will ultimately play the role of our $M$ (see Sec. 3.1). 


Although we will not present a detailed account of the rich theory of ADM decompositions (which can be found in Refs. \cite{gour06,gour07}, for example), we note that the 4-dimensional line element, in adapted coordinates $(t, x^{i})$ where $t$ is such that each spatial $x^{i}$ is constant along its field lines, can be written as
\begin{equation} \label{eq:adm}
ds^2_{g} = -n^2 dt^2 + h_{ij} \left(dx^{i} + \beta^{i} dt \right) \left( dx^{j} + \beta^{j} dt \right),
\end{equation}
where $n$ and $\boldsymbol{\beta}$ are known as the lapse function and shift vectors, respectively, and $h$ defines a Riemannian metric on each $\Sigma_{t}$. The metric $h$ is just the restriction of $g$ to $\Sigma_{t} \subset \mathcal{M}$, and will serve as the natural Riemannian metric uniquely associated to any given BH. 

\subsection{The Riemannian submanifold}

A BH is defined by the existence of a horizon $\mathcal{H}$ \cite{chand1}, the interior of which we do not include as part of the spacetime $\mathcal{M}$. For a given (stationary) BH geometry, the spatial slice $\Sigma_{0}$ can be identified as $\Sigma_{0} \simeq \mathbb{R}^3 / \mathcal{B}(\rH)$, where $\mathcal{B}(\rH)$ is a ball of radius $\rH$ centered around some origin for horizon radius $\rH$; i.e., one exorcises the interior. This construction suggests a natural candidate for $M$ for any given BH spacetime.

\begin{figure}[h]
\includegraphics[width=\textwidth]{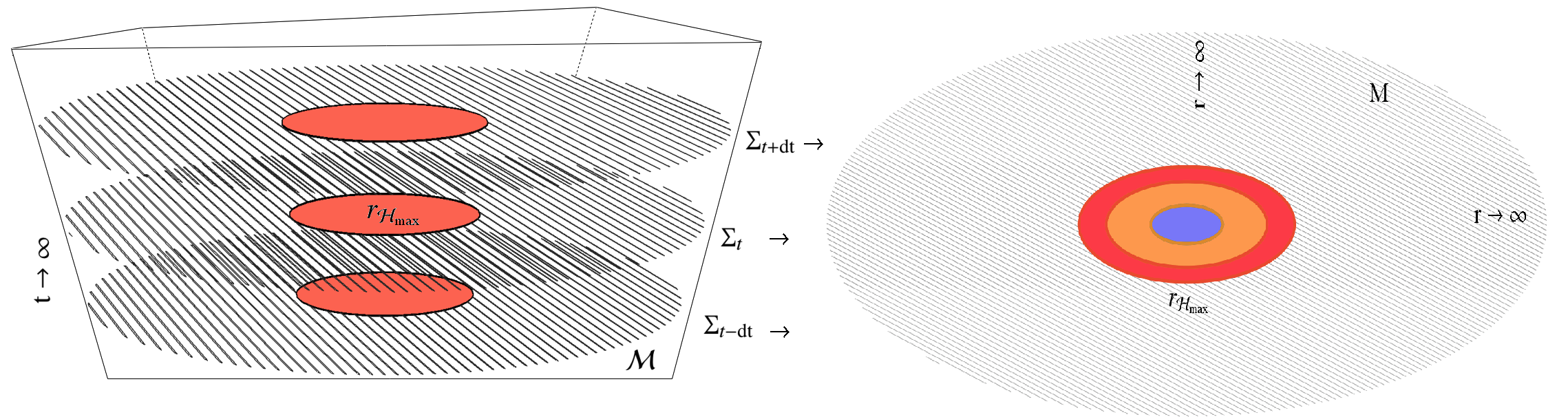}
\caption{Schematic diagram of a $3+1$ decomposition and the choosing of a suitable spacelike hypersurface $M$. For a stationary spacetime $\mathcal{M}$, we need only consider one leaf $\Sigma_{t}$ of the foliation, and the BH with the largest permitted horizon radius $r_{\mathcal{H}_{\text{max}}}$ among all $\mathcal{M}$ within the set under consideration defines $M$ through $M \simeq \mathbb{R}^3 / \mathcal{B}(r_{\mathcal{H}_{\text{max}}})$, i.e. $M$ is defined as the region between the outer edge of the largest horizon (illustrated by the red disc) and infinity. Sections of smaller BHs (those with horizons defined by the inner blue and orange discs) are still definable on $M$. \label{fig:fig0}}
\end{figure}

However, the volume of $\mathcal{H}$ is invariably a function of the hairs $\boldsymbol{q}$ (e.g. for the Schwarzschild metric of mass $m$, $\rH = 2 m$), which complicates the procedure of choosing a single $M$ from a family of BH spacetimes. One way to resolve this problem involves exploiting the diffeomorphism invariance and introducing a `horizon-adapted' radial coordinate $\tilde{r} = r / \rH(\boldsymbol{q})$ (say), so that the horizon is always located at $\tilde{r} = 1$ and all $M$ choices agree. Physically speaking, however, it is reasonable to consider only BHs with finite horizon area, so that there always exists some largest such $\rH$, $r_{\mathcal{H}_{\text{max}}}$, among any given set of BHs. For the Kerr case, this would correspond to setting a (arbitrarily large) maximum mass $m_{\text{max}}$ that any given BH could attain, and defining $M = \Sigma_{0} \simeq \mathbb{R}^3 / \mathcal{B}(2 m_{\text{max}})$. Spatial slices of spacetimes containing smaller BHs can still be defined on this particular $M$ in a consistent way through a restriction mapping. For concreteness, we adopt the latter approach here. A schematic representation of how $M$ is defined from some `maximal' BH spacetime $\mathcal{M}$ is shown in Figure \ref{fig:fig0} (see also Figure 7.1 in Ref. \cite{gour07}).

To pick an $m_{\text{max}}$, one could note that H$\beta$ emission-line measurements of the ultraluminous quasar TON 618 reveal the existence of a BH with mass $m \approx 6.6 \times 10^{10} M_{\odot}$ \cite{shem04}, which is the current record holder for heaviest known object. This suggests that $m_{\text{max}} \lesssim 10^{11} M_{\odot}$ would be a suitable choice, for example, though the geometric picture discussed in this paper is largely unaffected by the exact, numerical value of $m_{\text{max}}$ since  arbitrary units of mass can be chosen.

\subsection{Ensuring a finite superspace metric}

Although we have described a means to construct a universal Riemannian manifold $M$ from a given set of BH spacetimes, the construction of $\text{Met}_{\text{BH}}(M)$ cannot be completed unless the superspace metric $G$ is well defined. This requires the finiteness of the integral within expression \eqref{eq:metten}.

The Boyer-Lindquist coordinates traditionally used to describe BH spacetimes are singular at the horizon $\mathcal{H}$ \cite{chand1}, and $g^{-1}$ does not exist there. To remedy this, we introduce ingoing null coordinates $(V, r, \theta, \phi)$ (also known as Eddington-Finkelstein or Kerr coordinates) \cite{gour06} which define a principal null congruence (i.e. curves of constant $V$, constant $\theta$, and constant $\phi$ are null geodesics). An explicit time coordinate $t$ [as needed to build \eqref{eq:adm}] is then reintroduced through $V = t + r$. The metric $g$ (and hence $h$), when expressed in these ingoing null coordinates, is finite over the entirety of $\mathcal{M}$ (see Secs. 3.3 and 4.1 for some explicit constructions).

There is, however, an additional problem at infinity. Even for asymptotically flat spacetimes, the volume of $M$ is unbounded, and the integral defining $G$ (going from $r_{\mathcal{H}_{\text{max}}}$ to infinity) within \eqref{eq:metten} diverges. This issue may be overcome by introducing a conformal factor $\psi$ into the spacetime metric, which decays fast enough with $r$ to ensure the finiteness of the integral \eqref{eq:metten} over $M$. As such, we introduce a positive function $\psi$ through $\tilde{g}_{\alpha \beta} = \psi g_{\alpha \beta}$, with the property that $\underset{r \rightarrow \infty}{\lim} \psi = 0$. This defines a conformal 3-metric $\tilde{h}_{ij} = \psi h_{ij}$ from \eqref{eq:adm}. Importantly, conformal transformations do not disrupt the null congruence detailed above, since null geodesics for conformally related metrics are identical. In practice, the choice $\psi = 1/r$ is sufficient, though some investigation suggests that other choices do not introduce any qualitative differences.

\subsection{Static black holes}

For the case of static, spherically symmetric BHs, we carry out the above procedure in full with an arbitrary metric for demonstrative purposes. In this case, a general spacetime line element, in Boyer-Lindquist coordinates $(T,r,\theta,\phi)$, may be written as
\begin{equation} \label{eq:statmet}
ds^2_{g} = -A(r)^2 d T^2 + B(r)^2 dr^2 + r^2 d \theta^2 + r^2 \sin^2\theta d \phi^2,
\end{equation}
where the vanishing of $g^{rr}$ defines the location of the event horizon [i.e. $\rH$ is determined through $1/B(\rH) = 0$]. We begin by introducing the null coordinate $V$ through the transformation $dT \rightarrow dV - \left[ B(r)/A(r) \right] dr$, where the radial term within the parentheses is sometimes called the tortoise function. An explicit time coordinate $t$ is then reintroduced through $V = t + r$. In these latter coordinates $(t,r,\theta,\phi)$, the line element \eqref{eq:statmet} becomes
\begin{equation} \label{eq:nullmet}
ds^2_{g_{\text{null}}} = - A^2 dt^2 - 2 A \left( A - B \right) dt dr + A \left( 2 B - A \right) d r^2 + r^2 d\theta^2 +  r^2 \sin^2\theta d \phi^2,
\end{equation} 
which is more complicated than \eqref{eq:statmet} because of the off-diagonal components, though is well behaved at the horizon $\rH$ if the product $A(r) B(r)$ is finite there. The line element on $M$ is then found by directly comparing \eqref{eq:nullmet} with \eqref{eq:adm}, viz.
\begin{equation} \label{eq:conflin}
ds^2_{h} = A \left( 2 B - A \right) dr^2 + r^2 d \theta^2 + r^2 \sin^2 \theta d \phi^2.
\end{equation}
Finally, a conformal factor $\psi$ is introduced by rescaling $g$, leading to $ds^2_{\tilde{h}} = \psi ds^2_{h}$. Putting everything together, the configuration metric components $G_{ij}$ can be read off from \eqref{eq:metten},
\begin{equation} \label{eq:statconfig}
\hspace{-0.3cm}G_{ij} = 16 \pi \int^{\infty}_{r_{\mathcal{H}_{\text{max}}}} dr \frac {r^2 \psi(r)^{3/2}} {\left( 2 A B - A^2 \right)^{3/2}} \left[ \left( A - B \right) \frac {\partial A} {\partial q^{i}} - A \frac {\partial B} {\partial q^{i}} \right] \left[ \left( A - B \right) \frac {\partial A} {\partial q^{j}} - A \frac {\partial B} {\partial q^{j}} \right].
\end{equation}
Although relatively simple, the components \eqref{eq:statconfig} typically require numerical evaluation given some functions $A$ and $B$ and BH hairs $\boldsymbol{q}$. 

We close this section by noting that for static spacetimes, an alternate procedure to define a BH superspace could be through introducing Wick rotations $t \rightarrow i \tau$, as is done in Euclidean quantum gravity schemes \cite{eucqg}. This would remove the need to introduce ADM splits completely, and one could calculate distances using an appropriately restricted version of \eqref{eq:gilmed} defined over the (now) Riemannian 4-space with a Wick-rotated metric. Such an approach will be investigated elsewhere.


\section{Illustrative examples}

Here we present some explicit examples of the mathematical machinery developed in the previous sections. We focus on the case of vacuum GR (i.e. Kerr superspace; Sec 4.1), and on KZ static spacetimes (Sec. 4.2). 

\subsection{General relativity: the Kerr metric}

As is well-known, the Kerr metric uniquely represents stable, asymptotically flat BHs in vacuum GR \cite{mars1}. The Kerr metric has only two hairs, namely the mass $m$ and spin $a$, and so the configuration manifold of BHs in GR is rather simple. Expressed using the ingoing null (Kerr) coordinates $(t,r,\theta,\phi)$ described in Sec. 3.2 (see Appendix D of Ref. \cite{gour06} for how to relate these to the usual Boyer-Lindquist coordinates), the Kerr line element may be written as \cite{kerr63}
\begin{equation}
\begin{aligned}
ds_{\text{Kerr}}^2  =& - \left( 1 - \frac{2mr}{\rho^2} \right) dt^2 + \frac{4mr}{\rho^2} dt dr - \frac{4amr}{\rho^2} \sin^2\theta dt d\phi  \\
& + \left( 1 + \frac{2mr}{\rho^2} \right)  dr^2- 2 a \sin^2\theta \left( 1+ \frac{2mr}{\rho^2} \right) dr d\phi \label{eq:kerr} \\
& + \rho^2 d\theta^2 + \left( r^2 + a^2 + \frac{2 a^2 m r\sin^2\theta} {\rho^2} \right) \sin^2\theta d\phi^2,
\end{aligned}
\end{equation}
where $\rho^2 = r^2 + a^2 \cos^2 \theta$. The (outer) horizon $\rH$ is located at
\begin{equation} \label{eq:kerrhorizon}
\rH = m + \sqrt{m^2-a^2} ,
\end{equation}
which is maximal for $a=0$, and we note that the metric \eqref{eq:kerr} is regular there in any case. Because the horizon surface \eqref{eq:kerrhorizon} ceases to be real for $a > m$, the Kerr (BH) configuration space is only defined for $a/m \leq 1$. From \eqref{eq:adm}, we can read off the 3-metric components,
\begin{equation} \label{equation:gammaKerr}
    h_{ij} = \left( \begin{array}{ccc}
    1 + \frac{2mr}{\rho^2} & 0\  & 
         - a \left( 1+ \frac{2mr}{\rho^2}\right) \sin^2\theta \\
    0 & \rho^2 & 0 \\
    - a \left( 1+ \frac{2mr}{\rho^2}\right) \sin^2\theta \   & 0 &
        \left(r^2 + a^2  + \frac{2 a^2 m r \sin^2 \theta} {\rho^2} \right) \sin^2\theta       
    \end{array} \right) ,
\end{equation}
which can be used to define the configuration metric \eqref{eq:metten} after a suitable conformal factor $\psi$ has been introduced. The superspace metric $G$ on $\text{Met}_{\text{GR}}(M)$ can now be evaluated from \eqref{eq:metten} where $q^{i} = (m,a)$. For example, we find that the `$mm$-component' of the metric reads
\begin{equation}
G_{mm} = 8 \pi \int^{\pi}_{0} d \theta \int^{\infty}_{2 m_{\text{max}}} d r   \left[ \frac {r^2 + a^2 \cos^2 \theta} {\left(r^2 + 2 m r + a^2 \cos^2 \theta \right)^{3}} \right]^{1/2} r^2 \sin \theta \psi(r)^{3/2},
\end{equation}
which can be evaluated using standard numerical techniques without much difficulty after picking some value for $m_{\text{max}}$. Similar expressions can be obtained for the $ma$- and $aa$-components for the 2-dimensional metric $G$.  The methodology presented above can be adapted to more general stationary metrics in a straightforward way, provided that an appropriate null congruence can be found.

\begin{figure}[h]
\centerline{\includegraphics[width=0.493\textwidth]{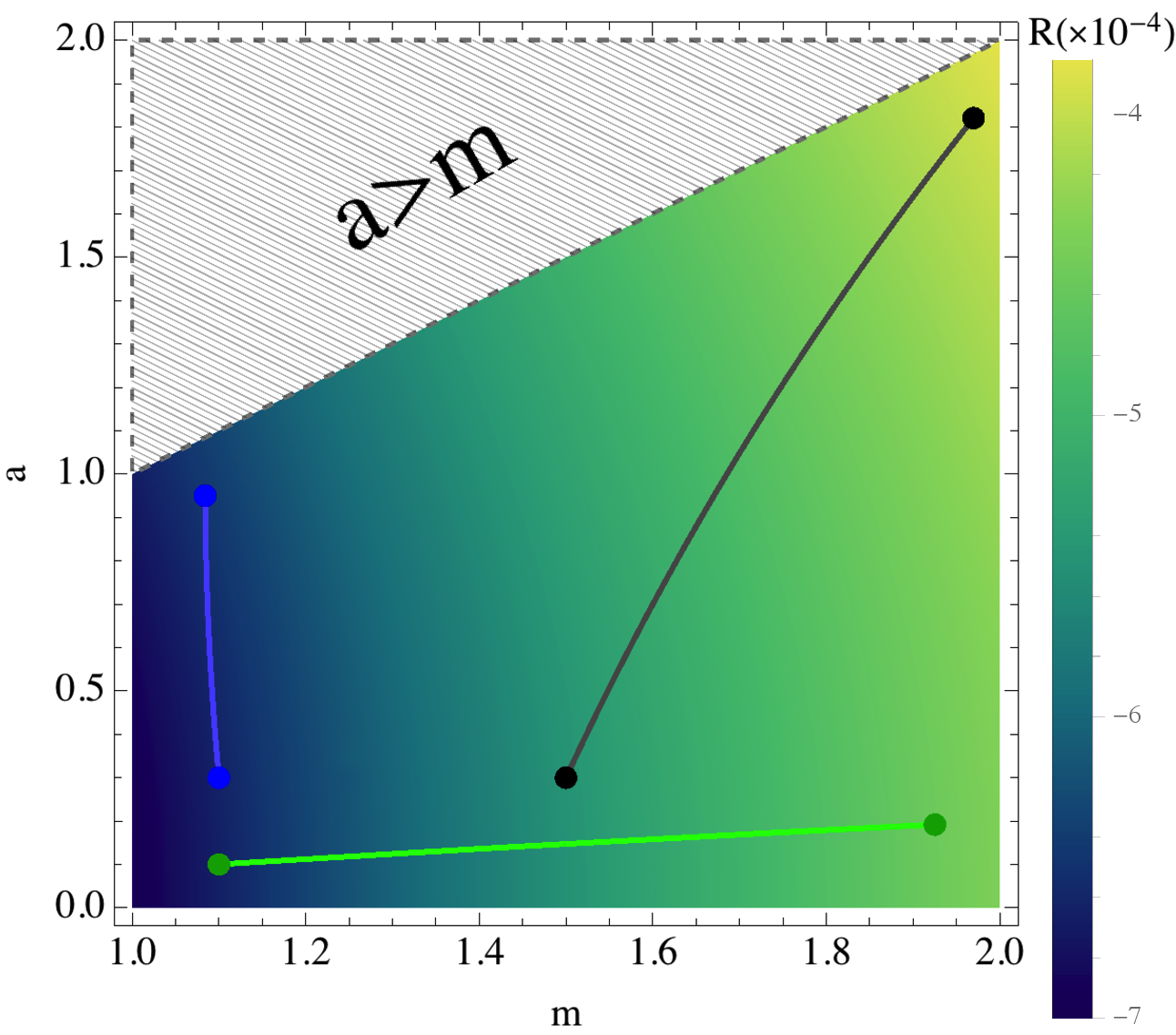}}
\caption{A sample of geodesic paths (solid blue, green, and black curves) on the $m \geq 1$ section of the configuration manifold of Kerr metrics with $m_{\text{max}} = 2$. The shaded region corresponds to $a > m$, which is disallowed by requring that the horizon \eqref{eq:kerrhorizon} be real. The colour scale, with lighter shades indicating a greater (more positive) value, shows the Ricci scalar $R$.  \label{fig:fig1}}
\end{figure}

In Figure \ref{fig:fig1} we show a variety of geodesics over the Kerr configuration manifold for $m_{\text{max}} = 2$ and $\psi(r) = 1/r$ with an overplotted colour scale that shows the scalar curvature $R$, as computed from \eqref{eq:ric}. The curvature $R$ is related to the geodesic distance since, in general, the ratio of volumes between a ball of radius $\delta \ll 1$ centered at some point $p$ on $\text{Met}_{\text{GR-BH}}(M)$ and a corresponding ball on the plane is proportional to the curvature $R$ at $p$ \cite{gray79}, i.e. \begin{equation} \label{eq:ricsc}
\frac{\text{Vol}[\mathcal{B}_{p}(\delta) \subset \text{Met}_{\text{GR-BH}}(M)]} {\text{Vol}[\mathcal{B}_{0}(\delta) \subset \mathbb{R}^{2}]} \approx 1 - \frac{R} {24} \delta^2.
\end{equation}
Although the curvature is relatively weak $(|R| \gtrsim 10^{-4})$ everywhere, we see that, by comparing the green [joining $(m=1.1,a=0.1)$ to $(m=1.93,a=0.19)$] and black [joining $(m=1.5,a=0.3)$ to $(m=1.97,a=1.82)$] paths (for example), geodesic curves deviate from straight lines more noticeably near the extremal boundary $a/m = 1$. In particular, as can be inferred from expression \eqref{eq:ricsc}, the superspace distance separating two near-extremal BHs (e.g. between ones with $a/m = 0.96$ and $a/m = 0.99$) is greater than the distance separating two slowly rotating holes (e.g. between ones with $a/m = 0.03$ and $a/m = 0.06$) even though the spin-value differences are the same. This agrees with the physical expectation that stronger spacetime distortions should accompany more rapidly rotating objects, and that a geometric distance measure should reflect this.

\subsection{Static, non-Einstein black holes: Konoplya and Zhidenko metric}

As discussed in the Introduction, finding exact solutions describing BHs in theories of gravity aiming to extend GR in some way can be a challenging task. However, progress can still be made in understanding how hypothetical, non-Einstein hairs may manifest in astrophysical data by studying the properties of some parameterised, non-Kerr metric. 

For example, a very general BH metric can be constructed by simply assuming Laurent expansions for the metric coefficients $g_{\mu \nu}$ with some arbitrary coupling parameters \cite{nkerr3}, which are to be constrained by observations. For the static case, one might write the metric potential $A$ appearing in \eqref{eq:statmet} as a series, $A = 1- \sum_{i = 1} a_{i} r^{-i}$ (or something similar), where only negative powers are kept to preserve asymptotic flatness. The $a_{i}$, which are to be constrained by observation, together with appropriate $b_{i}$ appearing in a series for the potential $B$, define the BH hairs. However, practical application usually requires that one truncate (or sum) the series, and so the question becomes: which terms should be kept? Recently, KZ \cite{kz20} showed that only a handful of terms [see expressions \eqref{eq:kz1} and \eqref{eq:kz2} below] are necessary to explore the quantitative features of these parameterised metrics for a variety of electromagnetic-based applications. Using the procedure outlined in Sec. 3.2, we can build the configuration manifold of KZ metrics to further test the relative importance of each $a_{i}$ and $b_{i}$, though in a purely geometric setting.

In particular, the KZ metric has four free parameters: $\rH$ (related to the BH mass; $r_{0}$ in the notation of KZ), $\epsilon$, $a_{1}$, and $b_{1}$, which appear within the metric in some particular combination. We can introduce an additional free parameter, $a_{2}$ (say), and see how this adjusts the structure of $\text{Met}_{\text{KZ-BH}}(M)$: if distances between metrics with large $a_{2}$ differences are comparatively small, for instance, this would imply that a non-zero $a_{2}$ has a weaker geometric impact than the KZ parameters. In particular, consider the metric \eqref{eq:statmet} with

\begin{equation} \label{eq:kz1}
A(r)^2 = 1 - \rH \left( \epsilon + 1 \right) / r + \rH^3 \left( \epsilon + a_{1} \right) / r^3 - \rH^4 \left( a_{1} + a_{2} \right) / r^4 + \rH^5 a_{2} / r^5,
\end{equation}
and
\begin{equation} \label{eq:kz2}
B(r)^2 =  A(r)^{-2} \left(1 + \rH^2 b_{1} /r^2 \right)^2,
\end{equation}
which reduces to the Schwarzschild metric in the limit $\epsilon = a_{i} = b_{i} = 0$, and the KZ metric \cite{kz20} for $a_{2} = 0$. We note that the product $AB$ is finite at $r=\rH$ so that \eqref{eq:conflin} is well-behaved, and that the absence of $r^{-2}$ terms in the coefficient $A$ ensures that the metric respects post-Newtonian constraints \cite{will18}. We may now evaluate the metric \eqref{eq:statconfig} where the $q^{i} = (\rH,\epsilon,a_{1},a_{2},b_{1})$ define coordinates on the 5-dimensional space $\text{Met}_{\text{KZ-BH}}(M)$.

\begin{figure}[h]
\centerline{\includegraphics[width=0.493\textwidth]{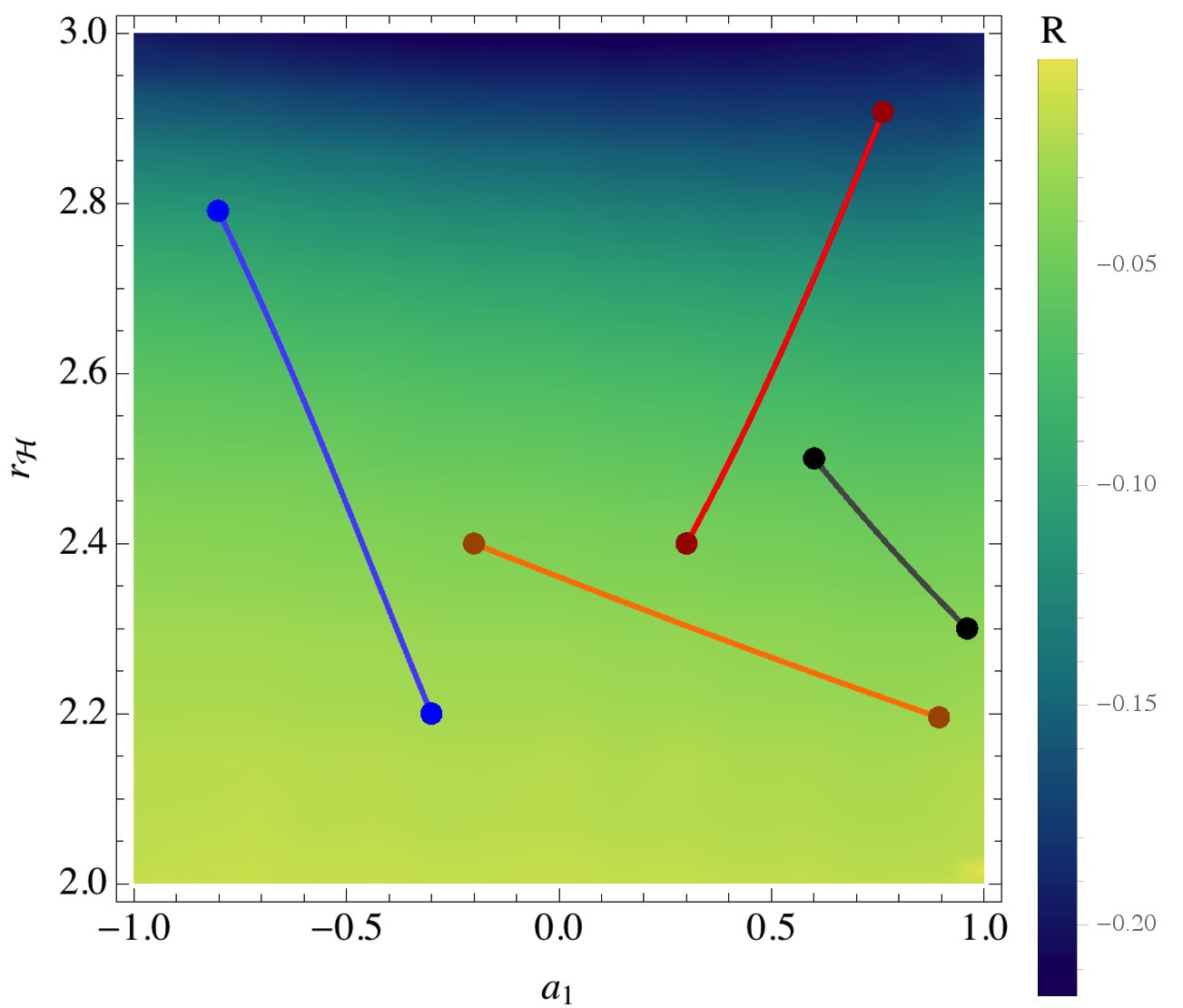}}
\caption{A sample of geodesic paths (solid blue, orange, red, and black curves) on the submanifold spanned by the $\rH$ and $a_{1}$ parameters within the space of KZ metrics. The colour scale, where brighter shades indicate a greater (more positive) value, shows the Ricci scalar $R$.  \label{fig:fig2}}
\end{figure}

In Figure \ref{fig:fig2} we show several geodesics (solid lines) together with the scalar curvature (colour scale) $R$, computed by contracting expression \eqref{eq:ric}, on a 2-dimensional slice of varying $a_{1}$ and $\rH$ with fixed $\epsilon = a_{2} = b_{1} = 0$, where we set $r_{\mathcal{H}_{\text{max}}} = 3$. The curvature is everywhere negative, and is greatest in magnitude for larger values of horizon radius; $\underset{|a_{1}| \leq 1,2 \leq \rH \leq 3}{\text{max}} |R| = 0.21$. This figure shows that the KZ configuration manifold admits, in general, a non-trivial geometric structure, and that bigger BHs with larger $\rH$ induce greater curvatures on the geometric superspace. Using the formula \eqref{eq:distance}, we find that (for example) the geodesic joining the points $(a_{1} = 0.6, \rH = 2.5)$ and $(a_{1} = 0.96, \rH = 2.3)$ (shown by the black curve) has length $d = 0.75$, while the flat distance $d_{\text{flat}} \equiv \sqrt{(2.5-2.3)^2 + (0.96-0.6)^2} = 0.41$.

\begin{figure}[h]
\centerline{\includegraphics[width=0.493\textwidth]{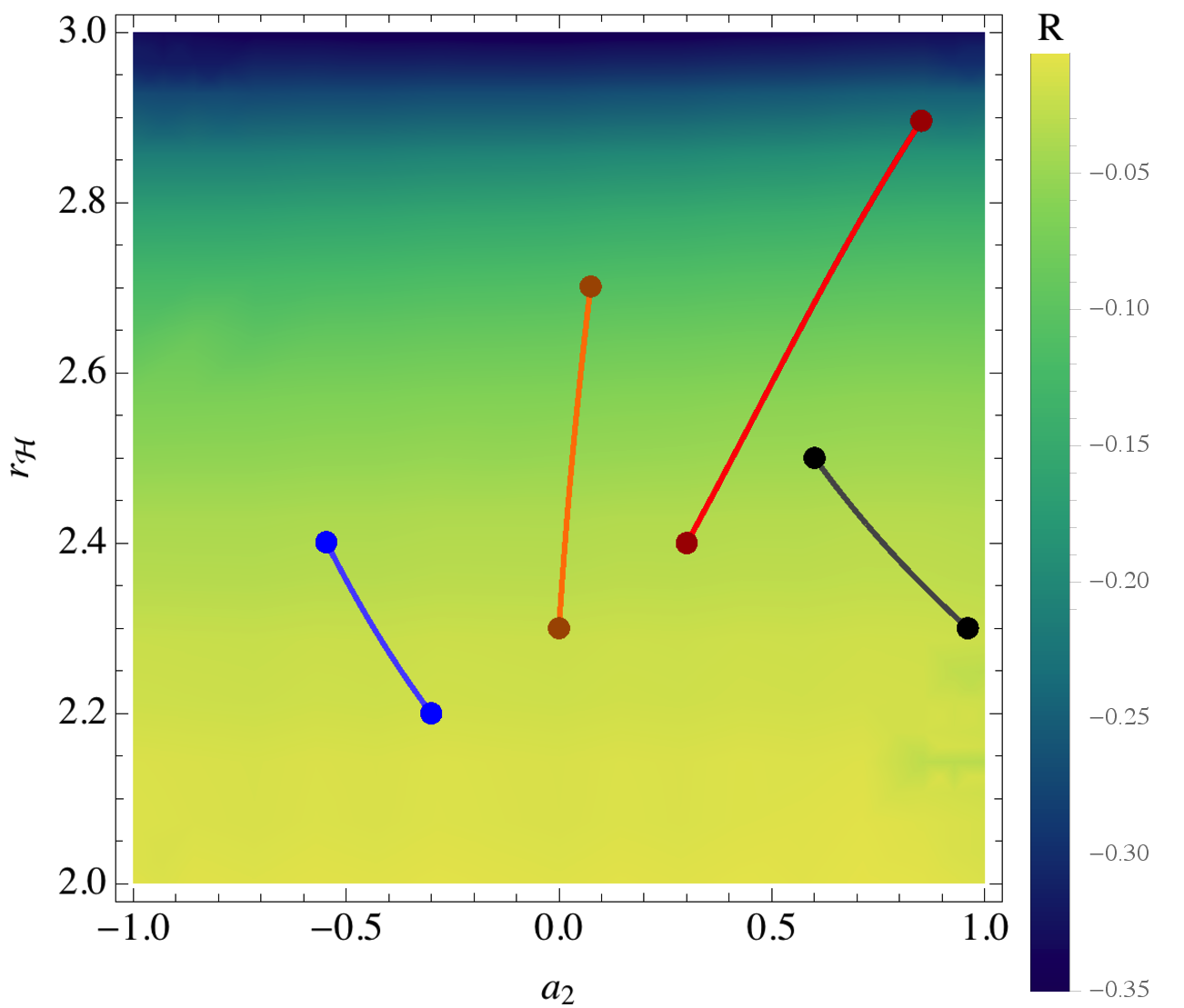}}
\caption{Similar to Fig. \ref{fig:fig2} though with fixed $a_{1} =0$ but varying $a_{2}$.  \label{fig:fig3}}
\end{figure}

For comparison, and to investigate the importance of the non-KZ parameter $a_{2}$, we show in Figure \ref{fig:fig3} a variety of geodesics and the scalar curvature on the slice with varying $\rH$ and $a_{2}$ but fixed $a_{1} = \epsilon = b_{1} = 0$. Overall, the respective scalar curvatures behave in a similar manner, though $R$ is generally more negative for the varying $a_{2}$ case, i.e. $\underset{|a_{2}| \leq 1,2 \leq \rH \leq 3}{\text{max}} |R| = 0.35$. 

We find that the distance between the points $(a_{2} = 0.6, \rH = 2.5)$ and $(a_{2} = 0.96, \rH = 2.3)$ (the geodesic connecting these points is shown by the black curve in Fig. \ref{fig:fig3}) is $d = 0.63$. These endpoints are the same for the black curve shown in Fig. \ref{fig:fig2}, though with $a_{2}$ replaced by $a_{1}$. This calculation reveals an important feature of the KZ metric: the distance between differing $a_{2}$ values is \emph{less} than the corresponding distance for different $a_{1}$ values, and therefore $a_{2}$ is less important, geometrically speaking, as metrics with different $a_{2}$ values are closer together. In short, when changing $a_{1}$ as opposed to $a_{2}$, the spaces are in a natural, geometric sense, more distinct. Although not shown, similar conclusions can be drawn by comparing other slices (e.g. the $\epsilon$-$a_{2}$ subspace) or other geodesics on the same slice. The superspace approach presented here therefore supports the conclusions given in Ref. \cite{kz20}: parameters other than $\epsilon$, $a_{1}$, $b_{1}$, and $\rH$ appear to matter less. 


\section{Some general remarks about the applicability of the superspace approach}

{It is important to stress that the localised superspace metric \eqref{eq:metten} is tailored to measure distances between various members of a parameterised family of metrics. Under these circumstances, distances \eqref{eq:distance} are relatively straightforward to compute and yield physically reasonable results, at least for the Kerr and KZ cases considered in Sec. 4. However, the general idea may break down when considering theories of gravity that admit separate families of BH solutions with mutually exclusive hairs. In this sense, there may not be an obvious parameterisation $h$ that interpolates between two disjoint solution branches. For example, there are no slowly rotating BHs that are simultaneously solutions in the Ho{\v{r}}ova-Lifshitz and Einstein-aether theories \cite{horlif}, even though the two are related in the infrared limit and admit the Schwarzschild solution as a limiting case. When considering sets of solutions with mutually exclusive hairs, the interpretation of distances on $\text{Met}_{\text{BH}}(M)$ as measuring BH similarity becomes less clear.}

{The superspace of \emph{metrics} also cannot account for non-metric hairs, and may thus fail to capture certain facets of `closeness'. As an extreme example, consider two scalar-tensor theories that differ only by the dynamics in their respective scalar sectors. These theories may admit BH solutions described by the same metric, but with different scalar field profiles (due to, e.g., spontaneous scalarization \cite{dam93,don18}). These two BH families are not to be thought of as identical, and yet the distance \eqref{eq:distance} between them on $\text{Met}_{\text{BH}}(M)$ for fixed metric parameters would be zero. In the same way, the metric \eqref{eq:metten} may also not reasonably quantify differences between solutions in bi- or multi-metric theories \cite{bimetric}.}


\section{Discussion}

Our ability to experimentally validate GR (or otherwise) in the strong-field regime is continually improving, with now over a dozen measurements of gravitational waves from events involving BHs \cite{abb}. For this reason, there is interest in developing generic, parameterised BH spacetimes \cite{nkerr1,nkerr2,nkerr3}, with the goal of using these to test for a variety of hypothetical deviations from the Kerr metric in a theory-agnostic way. On the other hand, a high level of generality implies that many of the proposed metrics can be mapped to one another \cite{nkerr5,nkerr4}, though it is not always obvious how this can be achieved in practice. It is therefore useful to have a tool which can quantitatively distinguish between various parameterised metrics, or indeed between non-Kerr metrics arising as exact solutions in beyond-Einstein theories of gravity. A similar approach for neutron star geometries was recently developed in Ref. \cite{suv20}.

Borrowing ideas from the study of geometrodynamics \cite{wheeler1,wheeler2,wheeler3}, we show in this paper how one can build a configuration manifold $\text{Met}_{\text{BH}}(M)$ of BHs; points on this non-flat manifold (see Figs. \ref{fig:fig1}--\ref{fig:fig3}) correspond to spatial $3$-metrics associated to ADM-split BH spacetimes. Distances on this manifold can be computed by solving the geodesic equations \eqref{eq:geodesic}. The construction of a geometric distance between two distinct metrics allows for a natural way to quantify the `closeness' of two BH geometries and could be used, for example, as a general measure for `non-Kerrness'; see also Ref. \cite{bhag18}. For the particular case of the Kerr configuration manifold (Sec. 4.1), which is 2-dimensional and is spanned by the mass $m$ and spin $a$, we found that the Ricci curvature, though remaining finite, develops stronger gradients as the extremal limit $a/m \rightarrow 1$ is approached (see Fig. \ref{fig:fig1}). The superspace approach therefore reaffirms the expectation that greater spacetime distortions accompany more rapidly rotating compact objects, as distances between BHs with greater $a$ values are larger than distances between slowly rotating holes.

For static BHs, we further validate the claims made by Konoplya and Zhidenko \cite{kz20} about which parameters are most important (that is, those which modulate astrophysical observables the most) when constructing a parameterised BH metric. This is achieved by noting that distances between (generalised) KZ metrics with an additional free parameter and those used in Ref. \cite{kz20} are comparatively small (see Sec. 4.2), which means that these spaces are geometrically similar. 

Although we have only considered asymptotically flat BHs here, it is likely that the approach can be generalised to the asymptotically de Sitter case without much difficulty, since the conformal factor $\psi$ can be used to tune the convergence of \eqref{eq:metten} even when $h$ grows as $r \rightarrow \infty$. The inclusion of non-stationary black holes, where time can be thought of as an additional hair of sorts, could also be achieved given a clear choice for a single $M$ among the family of metrics under consideration. {Finally, as discussed in Sec. 5, we note that the approach presented here may not capture all features that might be expected of BH `closeness'. Importantly, the localised metric \eqref{eq:metten} on the superspace $\text{Met}_{\text{BH}}(M)$ requires a parameterised family of BH metrics $h$. It is not obvious how to construct such a parameterisation for theories admitting disjoint branches of solutions with mutually exclusive hairs \cite{horlif,don18}. Purely non-metric charges are also not accounted for in expression \eqref{eq:metten}, as non-metric (e.g. scalar) fields do not appear.}

\section*{Acknowledgements}
We thank Prof. Bill Moran for hospitality shown during a research visit to Melbourne, where some of this work was completed. We are grateful to Sebastian V{\"o}lkel for discussions. {We thank the anonymous referees for providing feedback which helped improve the clarity of this manuscript.} This work was supported by the Alexander von Humboldt Foundation.


\section*{References}

\begin{thebibliography}{unsrt}




\bibitem{wheeler1}
J. A. Wheeler, \emph{Einsteins Vision},
Springer Verlag, Berlin, 1968

\bibitem{wheeler2}
J. A. Wheeler, 
in Battelle Recontres, Editors, DeWitt and Wheeler, W. A. Benjamin, Inc., 1968.

\bibitem{wheeler3}
J. A. Wheeler, 
Annals of Physics \textbf{2}, 604 (1957).

\bibitem{gil2}
B. S. DeWitt, 
Phys. Rev. \textbf{160}, 1113 (1967).

\bibitem{clark10}
B. Clarke, 
Calculus of Variations and Partial Differential Equations \textbf{39}, 533 (2010).

\bibitem{ebin}
D. G. Ebin, 
Proc. Symp. Pure Math. AMS \textbf{15}, 11 (1970).

\bibitem{gil1}
O. Gil-Medrano and P. W. Michor, 
The Quarterly Journal of Mathematics \textbf{42}, 183 (1991).



\bibitem{fishcer}
A. Fischer, \emph{The theory of superspace}, in Proceedings of the Relativity Conference, edited by M. Carmeli, S. I. Fickler, and L. Witten (Plenum, New York, 1970).

\bibitem{edwards}
D. A. Edwards, 
Studies in topology. Academic Press, 121, (1975).

\bibitem{guil09}
D. Giulini, 
Gen. Rel. Grav. \textbf{41}, 785 (2009).


\bibitem{hawk}
S. W. Hawking, 
Comm. Math. Phys. \textbf{55}, 133 (1977).

\bibitem{eucqg}
G. W. Gibbons and S. W. Hawking, \emph{Euclidean Quantum Gravity},
(World Scientific, Singapore, 1993).

\bibitem{kerr63}
R. P. Kerr, 
Phys. Rev. Lett. \textbf{11}, 237 (1963).

\bibitem{mars1}
M. Mars, 
Class. Quant. Grav. \textbf{17}, 3353 (2000).

\bibitem{chand1}
S. Chandrasekhar, \emph{The Mathematical Theory of Black Holes},
(Oxford University Press, Oxford, 1998).

\bibitem{schconfig}
K. V. Kucha{\u{r}}, 
Phys. Rev. D \textbf{50}, 3961 (1994).

\bibitem{exact1}
D. Grumiller and N. Yunes, 
Phys. Rev. D \textbf{77}, 044015 (2008).

\bibitem{exact2}
T. Clifton and J. D. Barrow, 
Phys. Rev. D \textbf{72}, 103005 (2005).

\bibitem{exact3}
C. Charmousis et al., 
Phys. Rev. D \textbf{100}, 084020 (2019).

\bibitem{exact4}
R. A. Konoplya and A. Zhidenko, 
Phys. Rev. D \textbf{101}, 084038 (2020).



\bibitem{bhag18}
S. Bhagwat et al., 
Phys. Rev. D \textbf{97}, 104065 (2018).

\bibitem{nkerr1}
K. Glampedakis and S. Babak, 
Class. Quant. Grav. \textbf{23}, 4167 (2006).

\bibitem{nkerr2}
S. Vigeland, N. Yunes, and L. C. Stein, 
Phys. Rev. D \textbf{83}, 104027 (2011).

\bibitem{nkerr3}
T. Johannsen and D. Psaltis, 
Phys. Rev. D \textbf{83}, 124015 (2011).

\bibitem{contfrac}
L. Rezzolla and A. Zhidenko, 
Phys. Rev. D \textbf{90}, 084009 (2014).

\bibitem{nkerr5}
G. O. Papadopoulos and K. D. Kokkotas, 
Class. Quant. Grav. \textbf{35}, 185014 (2018).

\bibitem{nkerr4}
Z. Carson and K. Yagi, 
Phys. Rev. D \textbf{101}, 084030 (2020).


\bibitem{kz20}
R. A. Konoplya and A. Zhidenko, 
Phys. Rev. D \textbf{101}, 124004 (2020).

\bibitem{osmo}
O. Pekonen, 
J. Geom. Phys. \textbf{4}, 493 (1987).


\bibitem{pappas15}
G. Pappas and T. P. Sotiriou, 
Phys. Rev. D \textbf{91}, 044011 (2015).

\bibitem{suvmel16}
A. G. Suvorov and A. Melatos, 
Phys. Rev. D \textbf{93}, 024004 (2016).




\bibitem{ymbh1}
R. Bartnik and J. McKinnon, 
Phys. Rev. Lett. \textbf{61}, 141 (1988).

\bibitem{ymbh2}
P. Bizon, 
Phys. Rev. Lett. \textbf{64}, 2844 (1990).

\bibitem{suv20}
A. G. Suvorov, 
Class. Quant. Grav. \textbf{37}, 065013 (2020).



\bibitem{admsplit}
R. L. Arnowitt, S. D. Deser, and C. W. Misner, 
Gen. Rel. Grav. 40, 1997 (1962).

\bibitem{ger70}
R. Geroch, 
J. Math. Phys. \textbf{11}, 437 (1970).

\bibitem{gour06}
E. Gourgoulhon and J. L. Jaramillo, 
Physics Reports \textbf{423}, 159 (2006).

\bibitem{gour07}
E. Gourgoulhon, \emph{Lecture Notes in Physics},
(Springer Verlag, Berlin, 2012), Vol. 846

\bibitem{shem04}
O. Shemmer et al., 
Astrophys. J. \textbf{614}, 547 (2004).




\bibitem{gray79}
A. Gray and L. Vanhecke, 
Acta Mathematica \textbf{142}, 157 (1979).

\bibitem{will18}
C. M. Will, \emph{Theory and experiment in gravitational physics},
(Cambridge university press, Cambridge, 2018).

\bibitem{horlif}
E. Barausse and T. P. Sotiriou, 
Phys. Rev. D \textbf{87}, 087504 (2013).

\bibitem{dam93}
T. Damour and G. Esposito-Far{\'e}se, 
Phys. Rev. D \textbf{54}, 1474 (1996).

\bibitem{don18}
D. D. Doneva and S. S. Yazadjiev, 
Phys. Rev. Lett. \textbf{120}, 131103 (2018).

\bibitem{bimetric}
E. Babichev and R. Brito, 
Class. Quant. Grav. \textbf{32}, 154001 (2015).

\bibitem{abb}
B. P. Abbott et al., 
Phys. Rev. X \textbf{9}, 031040 (2019).





\end{thebibliography}
\end{document}